\documentclass[
 twocolumn,
 amsmath,amssymb,
 aps, prb,
]{revtex4-2}

\usepackage{graphicx}
\usepackage{dcolumn}
\usepackage{bm}
\usepackage{float}
\usepackage{caption}
\usepackage{subcaption}
\usepackage{ragged2e}
\usepackage{xcolor}  % just to write some text in red but you can remove afterwards
\usepackage[colorlinks=true, citecolor=blue, linkcolor=blue, urlcolor=blue]{hyperref}

\begin{document}

\preprint{APS/PRB}

\title{\textbf{Superconductivity and electronic structure evolution in the enforced semimetal Fe-doped ZrTe$_2$} 
}% 

\author{L. M. Ishikura}
\email{larissaishiikura@usp.br}
\affiliation{Escola de Engenharia de Lorena - DEMAR, Universidade de São Paulo, 12612-550, Lorena, Brazil}

\author{C. F. Schuch}
\email{caua.schuch@usp.br}
\affiliation{Escola de Engenharia de Lorena - DEMAR, Universidade de São Paulo, 12612-550, Lorena, Brazil}

\author{F. F. Nogueira}
\affiliation{Escola de Engenharia de Lorena - DEMAR, Universidade de São Paulo, 12612-550, Lorena, Brazil}

\author{L. E. Corrêa}
\affiliation{Escola de Engenharia de Lorena - DEMAR, Universidade de São Paulo, 12612-550, Lorena, Brazil}

\author{L. R. de Faria}
\affiliation{Escola de Engenharia de Lorena - DEMAR, Universidade de São Paulo, 12612-550, Lorena, Brazil}

\author{A. Fa\'{e} Rabello}
\affiliation{Laboratory for Quantum Matter under Extreme Conditions,
Instituto de Física, Universidade de São Paulo, 05508-090, São Paulo, Brazil}

\author{J. Larrea Jim\'{e}nez}
\affiliation{Laboratory for Quantum Matter under Extreme Conditions,
Instituto de Física, Universidade de São Paulo, 05508-090, São Paulo, Brazil}

\author{L. T. F. Eleno}
\affiliation{Escola de Engenharia de Lorena - DEMAR, Universidade de São Paulo, 12612-550, Lorena, Brazil}

\author{A. J. S. Machado}
\email{ajefferson@usp.br}
\affiliation{Escola de Engenharia de Lorena - DEMAR, Universidade de São Paulo, 12612-550, Lorena, Brazil}

\date{\today}% It is always \today, today,

\begin{abstract}

ZrTe$_2$ is an outstanding layered semimetal due to the topologically nontrivial electronic structure. In this work, we present an investigation of the electronic evolution of ZrTe$_2$ in the presence of Fe intercalation, namely Fe$_{x}$ZrTe$_2$ ($x= 0 - 0.25$), scrutinized by both experimental measurements and \textit{ab} initio calculations. While the first reveals a superconducting state with a maximum critical temperature $T_c = 2.74$ K ($x=$ 0.03), the latter indicates that the topological features of the pristine ZrTe$_2$ is sensitive to the distance between Te atoms and Zr layers. Also, the intercalation of Fe does not modify the non-trivial electronic band structure unlike the band crossings are now shifted slightly below $E_{F}$. In particular, a van Hove singularity near the Fermi level for a Fe content of $x=0.125$ is observed in the density of states, indicating that the superconducting order may be associated with features of the unfolded band structure and the concomitant enhancement of the density of states at $E_F$. Finally, our results reveal that the new compound with inclusion of Fe intercalation preserves the enforced semimetal classification.

\end{abstract}

\maketitle

%\tableofcontents

\section{Introduction}

Transition metal dichalcogenides (TMDs) are layered materials with the general chemical formula MX$_2$, where M denotes a transition metal and X a chalcogen element (S, Se, or Te) \cite{manzeli2017,wang2012,chhowalla2013}. Their crystal structure consists of stacked X–M–X layers held together by weak van der Waals (vdW) interactions to form bulk crystals. TMDs exhibit a variety of coherent states and electrical instabilities, including superconductivity and charge density waves (CDWs) \cite{lian2023,ledneva2022,zhou2016,han2018,bhoi2016}. In addition, recent studies have reported the emergence of nontrivial topological states in several TMDs, particularly type-II Dirac fermions \cite{bahramy2018,ferreira2021,fei2017,kar2020,belopolski2016}.

The presence of vdW gaps between layers enables the intercalation of atoms and molecules, which has been shown to significantly modify electronic properties of the host material \cite{wagner2008,morosan2010,kiswandhi2013,chang2016}. This work focuses on ZrTe$_2$ TMD (M = Zr and X = Te), which crystallizes in the prototype CdI$_2$ structure \textit{P-3m1} (space group 164) \cite{shkvarina2018}. Recent results of angle-resolved photon-emission spectroscopy (ARPES) and de Haas-van Alphen oscillation experiments suggest that this compound can be classified as a Dirac semimetal with 4-fold massless quasiparticles \cite{tsipas2018,nguyen2022}. Ionic intercalation in ZrTe$_2$ may lead to superconductivity, as reported in Cu$_x$ZrTe$_2$, with critical temperatures up to $T_c \approx 9.0$ K \cite{machado2017}, in Pt$_x$ZrTe$_2$ with $T_c\approx 3.5$ K \cite{correa2024}. Last but not least, the coexistence of superconductivity with charge density wave (CDW) order was reported in Ni$_x$ZrTe$_2$, with $T_{c} \approx 4.0$ K and $T_{CDW} \approx 287.0$ K \cite{correa2022}.

Here, we address the effect of Fe intercalation in ZrTe$_2$ electronic properties, namely Fe$_{x}$ZrTe$_{2}$ single crystals. Our investigation of the electronic properties was conducted by electrical resistivity and magnetoresistance combined with first-principle $ab$ initio calculations of the electronic-structure for the ideal limit $x=1$ and for disordered doped compositions $x=0.125,0.25$, using supercells. Our experimental results show that a superconducting state with multi-band characteristic behavior emerges in coexistence with CDW order. In DFT part, multiple pristine ZrTe$_2$ configurations were tested, to address the divergence in the literature where both NLSM and DSM were reported. The density of states calculation using supercell for the lower $x$ structure revealed a van Hove singularity near the Fermi level, a system instability that could justify the emergence of the superconducting order. Finally, we also analyzed the topological features in the bandstructure of Fe$_{x}$ZrTe$_2$ by tracking its symmetry eigenvalues and traces for all operators to employ the Topological Quantum Chemistry (TQC) workflow, which gave the result that it still keeps the same enforced semimetal character, with a crossing between $\Gamma-A$, plus a new one in $H-K$ at the Brillouin zone boundary.

\section{Experimental procedure}

Initially, polycrystalline samples with Fe$_x$ZrTe$_2$ nominal composition, with $x = 0.125, 0.25$, were prepared by reacting the stoichiometric amount of high purity Zr and Fe sheets with Te chunks at 950°C in a sealed quartz tube with 150 torr of UHP argon gas for 48 h, which was later quenched in water. About 0.5 g of each sample was then ground and pressed into 8 mm diameter pellets, sealed in quartz tubes under argon and heat treated at the same temperature for an additional 48 h. For Fe$_x$ZrTe$_2$ single crystal growth, the isothermal chemical vapor transport (ICVT) methodology was used \cite{correa2022icvt}. Regarding this, the polycrystalline pellets were firstly quenched, reground and pressed again, and sealed in quartz tubes now under vacuum with the addition of 30 mg of I$_2$, which were placed horizontally in box furnaces at 1000°C for 7 days and lastly quenched in water. Plate-like single crystals as big as 10 × 10 mm$^2$ grew out from the pellet and were gently removed with a pair of tweezers.
The crystal structure and effects of intercalation on cell parameters were studied by x-ray diffraction (XRD) using a Malvern Panalytical Empyrean diffractometer equipped with a texture goniometer. The actual composition of the crystals was determined from energy dispersive microscopy (EDS) in a Hitachi TM 3000 scanning electron microscope (SEM). We found a deviation from the nominal concentration $x$, with $\sim$12.5 \% of the original value being incorporated into the single crystal. 
Measurements of electrical transport, i.e., electrical resistance and magnenoresistance in four-probe configuration, were carried out in an Evercool II Quantum Design Physical Properties Measurement System (PPMS) in a temperature range between 1.9 K and 300 K and under applied magnetic fields up to 0.2 T. Copper wires of 0.1 mm diameter were attached to the surface of the crystal using Ted Pella silver paint.

\section{Computational methods}
\begin{figure}
    \centering

    \begin{subfigure}[b]{0.51\linewidth}
        \centering
        \includegraphics[width=\linewidth]{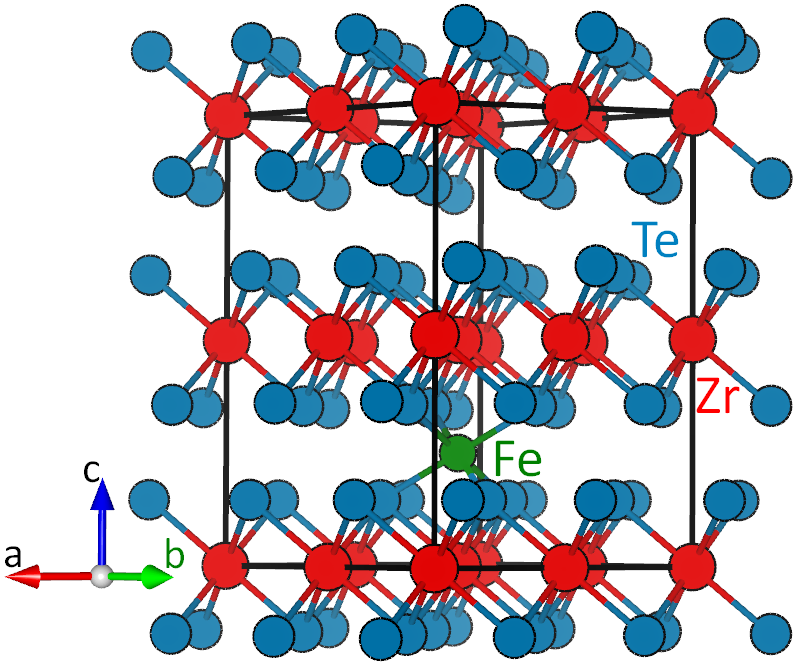}
        \caption{}
        \label{fig:sg164data:a}
    \end{subfigure}
    \hfill
    \begin{subfigure}[b]{0.43\linewidth}
        \centering
        \includegraphics[width=\linewidth]{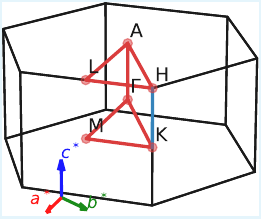}
        \caption{}
        \label{fig:sg164data:b}
    \end{subfigure}

    \caption{\justifying
    (a) Crystal structure of Fe-doped ZrTe$_2$ in the $2\times2\times2$ supercell employed in DFT calculations, representing the octahedral intercalation site. (b) 1st Brillouin Zone of SG 164 with highlighted path (red) used to plot the dispersion curves, and extra path $H-K$ (blue) where the other band crossing occurs.}
    \label{fig:sg164data}
\end{figure}

Ab initio electronic-structure calculations were performed using the plane-wave
package \textsc{Quantum ESPRESSO} (QE) \cite{giannozzi2009,Giannozzi_2017} within the
Kohn--Sham density functional theory (DFT) formalism \cite{kohn1965}. Structural relaxations
were carried out using the next-generation \textsc{vdW--DF3} exchange--correlation
functional \cite{vdw-d3,vdw2,vdw-review1,vdw-review2}, neglecting spin--orbit coupling
(SOC), in order to properly account for the nonlocal van der Waals (vdW) interactions between the
ZrTe$_2$ layers, while subsequent electronic-structure calculations were performed using
the generalized gradient approximation (GGA) in the Perdew--Burke--Ernzerhof (PBE)
parametrization \cite{gga-pbe}, with the projector augmented wave (PAW) method \cite{PhysRevB.50.17953} employed to
describe the interaction between valence electrons and ionic cores. The pseudopotentials (with SOC) used
were generated using the \texttt{ld1.x} atomic code of QE, based on the \textsc{PSLIBRARY}
inputs (version~$1.0.0$) \cite{PSlibrary}. We also employed the \textsc{Supercell} program \cite{Okhotnikov2016Supercell} to build $2\times2\times2$ supercells (SC), to obtain a better description of the intercalation effects, following the same schema for structural relaxations as mentioned before. The kinetic-energy cutoff for the plane-wave basis
was fully converged at $110$~Ry, with a charge-density cutoff of $8\times110$~Ry, and a
Marzari--Vanderbilt--De~Vita--Payne cold smearing of $0.01$~Ry was used
\cite{marzari1999thermal}. The convergence thresholds in relaxations for total energies and atomic forces
were set to $10^{-6}$~Ry and $10^{-5}$~Ry/$a_0$ for the primitive cell, and to $10^{-4}$~Ry
and $10^{-3}$~Ry/$a_0$ for the SC calculations, respectively, and for the electronic selfconsistency set to $10^{-10}$~Ry for all calculations.
Self-consistent field (SCF) calculations employed $\Gamma$-centered Monkhorst--Pack
$k$-point meshes of $16\times16\times8$ for the primitive cell and $8\times8\times8$ for the
SC \cite{monkhorst_pack_1976}, while non-self-consistent field (NSCF) calculations
used denser meshes of $20\times20\times12$ for the primitive cell and $10\times10\times8$ for SC; the density of states
(DOS) was computed using the tetrahedron method in Bl\"ochl’s implementation
\cite{PhysRevB.49.16223}. Electronic band structures were calculated along the
high-symmetry path $\Gamma$--$M$--$K$--$\Gamma$--$A$--$L$--$H$--$A$, following the convention
of Setyawan et al \cite{setyawan2010}, as shown in Fig.~\ref{fig:sg164data}(b). To accurately capture the
three-dimensional bulk effects of Fe intercalation within the SC approach and the
resulting two-component spinor eigenfunctions, band unfolding calculations were performed
using the \textsc{Ban Duppy} code \cite{unfold1,unfold2,IrRep}, along the $A$--$\Gamma$--$M$ path
for Fe$_x$ZrTe$_2$ ($x = 0.0$, 0.125, and 0.25), based on the projection of SC bands spectral weights
onto the primitive Brillouin zone. Symmetry analysis of the electronic eigenstates and irreducible representations (irreps) calculation were performed with \textsc{IrRep} \cite{IrRep}, and its outputs containing the simmetry operation little group traces were used into Check Topological Mat. tool \cite{doi:10.1126/science.abg9094,catalogue-topo} from BCS, in order to diagnose the compounds topology, based in the Topological Quantum Chemistry advances \cite{TQC}. Fermi
surface and electron localization function visualization were made using \textsc{FermiSurfer} \cite{KAWAMURA2019197} and \textsc{VESTA} \cite{VESTA} codes,
respectively.

\section{Results and discussion}

\subsection{Crystal structure}

ZrTe$_2$ compound crystalizes in a centrosymmetric low dimensional (2D)
trigonal structure with symmorphic
space group P$\bar{3}$m$1$ (SG No. $164$), CdI$_2$ prototype. In this crystal, Zr atoms occupy the cell corners $1a$ Wyckoff site, in the center of an octrahedral bonding environment with six equivalent Te atoms in the $2d$ site. Basically, the stacking of the $a-b$ planes occurs in $c$ direction, where Fe atoms in $1b$ site enters the structure in another octahedral center but dislocated in $0.5c$, closing the van der Waals gap. The other possibility of intercalation site in the gap is a tetrahedral position in $2d$ site in same $a-b$ plane coordinates as Te atoms. The 3D bulk ZrTe$_2$ has 12 symmetries operations (Seitz notation), including identity $\{1|0\}$ and inversion $\{-1|0\}$: the threefold rotation around $c$ direction $\{3^{\pm}_{001}|0\}$; the twofold rotations in $a/b$ direction $\{2_{100}|0\},\{2_{010}|0\}$ ($\{2_{110}|0\}$) and the reflection $\{m_{100}|0\},\{m_{010}|0\}$ ($\{m_{110}|0\}$) in respect to a plane of the family $(100)$ ($(110)$); the rotoinversion arounc $c$-axis $\{-3^{\pm}_{001}|0\}$, generating the stacked, low-dimensional crystal. The high-symmetry points of the BZ $\Gamma=(0,0,0)$, $A=(0,0,\frac{1}{2})$ have all the symmetries of the SG, and $K=(\frac{1}{3},\frac{1}{3},0)$ and $H=(\frac{1}{3},\frac{1}{3},\frac{1}{2})$, object of this study, do not have inversion, rotoinversion and reflection symmetries.

We further verified that the octahedral intercalation site is energetically more favorable than tetrahedral, by fully relaxing $x=0.125$ supercells with one Fe atom, in which the first has a lower formation enthalpy than the latter by $\Delta H=-8.362$ meV/atom. Also, a more in depth analysis of the Fe positions was explored by appyling Mossbauer results, which will be part in another publication, but those results are in convergence with our DFT data, ($95\%$ and $5\%$ for the two sites occupation in the crystal). Thus, all the subsequent calculations were perfomed in octahedral site. By comparison of experimental and ab initio lattice parameters, shown in Fig. \ref{fig:exp_drx}(c), we see that the vdW interactions are properly described by the functional employed, with an excellent agreement of less than $1\%$ difference between the measures and the DFT. Also, the contraction of $c$ parameter is well and continuous represented up to $x=1.0$ (see Table S1 of Supp. Material), and by performing a linear fit in DFT approach, the compositions of $x\approx0.05$ have $c\approx6.604$ \AA. The $a$ parameter is also expected to shrink through increasing $x$, but at higher Fe percentages it suffers an expansion to accomodate a full site occupation in $a-b$ plane (no more vdW interaction), this will be further explored in the next section.

\subsection{Experimental results}
\label{expt_res}
\begin{figure*}
    \includegraphics[width=\linewidth]{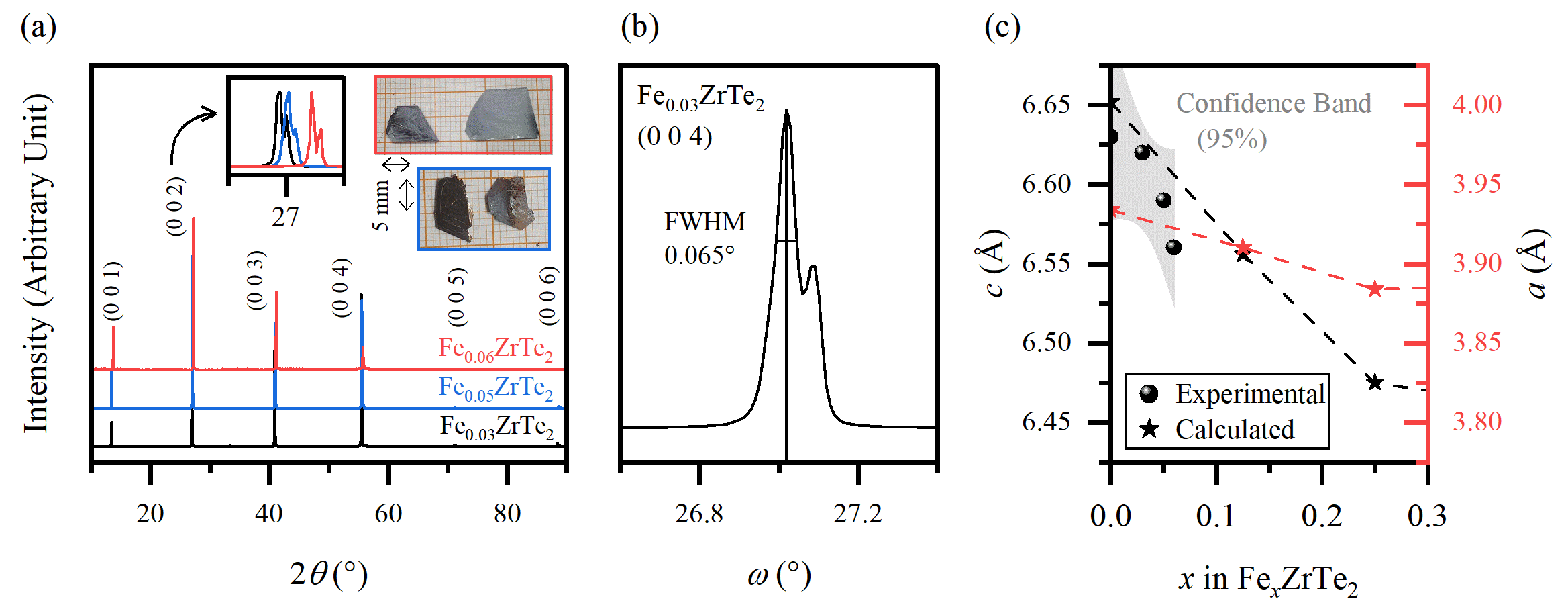}
    \caption{\justifying (a) $\theta-2\theta$ XRD diffractogram for Fe$_x$ZrTe$_2$ single crystals ($x=0.03,0.05,0.06$) revealing only $(00\ell)$ reflections. The insets show the $(002)$ peak shift between different Fe concentrations and pictures of single crystals grown via ICVT, with dimensions up to $\sim 10.0 \times10.0\times0.2$ mm$^3$; (b) Rocking curve centered at the $(004)$ peak of a Fe$_{0.03}$ZrTe$_2$ single crystal, with a FWHM = 0.065°; (c) Experimental and DFT calculated lattice parameters $c$ and $a$ as a function of $x$ in Fe$_x$ZrTe$_2$. Experimental $c$ data was obtained by fitting the $(00\ell)$ reflections from Fig. \ref{fig:exp_drx}(a).}
    \label{fig:exp_drx}
\end{figure*}

The X-ray diffractograms shown in Fig. \ref{fig:exp_drx}(a) represent a $\theta-2\theta$ scan of the flat facet obtained from the different composition synthesized Fe$_x$ZrTe$_2$ single crystals under Cu k$\alpha$ radiation. These crystals revealed exclusively $(00\ell)$ reflections, which correspond to the $a-b$ hexagonal planes of the ZrTe$_2$ structure. EDS composition analysis of grown crystals is also shown in Fig. \ref{fig:exp_drx}(a), indicating that approximately 12.5\% of Fe content in the precursor pallets was incorporated into the single crystals. Because the 1:2 atomic ratio between Zr and Te contents remains unaltered, Fe is likely intercalated in vdW gaps rather than substituting the Zr or Te atoms. Fig. \ref{fig:exp_drx}(b) shows the omega scan of the $(004)$ Fe$_{0.03}$ZrTe$_2$ peak, where the narrowed Full Width at Half Maximum (FWHM) of 0.065° indicates excellent quality of our single crystals. The shifting of the peaks to higher $2\theta$ values with increasing Fe concentration (see the inset of Fig. \ref{fig:exp_drx}(a)), is indicative of a decrease in the lattice parameter $c$ associated with the distance between the hexagonal planes in ZrTe$_2$ structure. The values of the $c$ parameter, roughly estimated by fitting only the $(00\ell)$ reflections in $\theta-2\theta$ scans, are plotted as a function of $x$ content in Fe$_x$ZrTe$_2$ in Fig. \ref{fig:exp_drx}(c). This figure also depicts the good agreement between our experimental lattice parameters and those obtained from our DFT calculations.

\begin{figure*}
    \includegraphics[width=\linewidth]{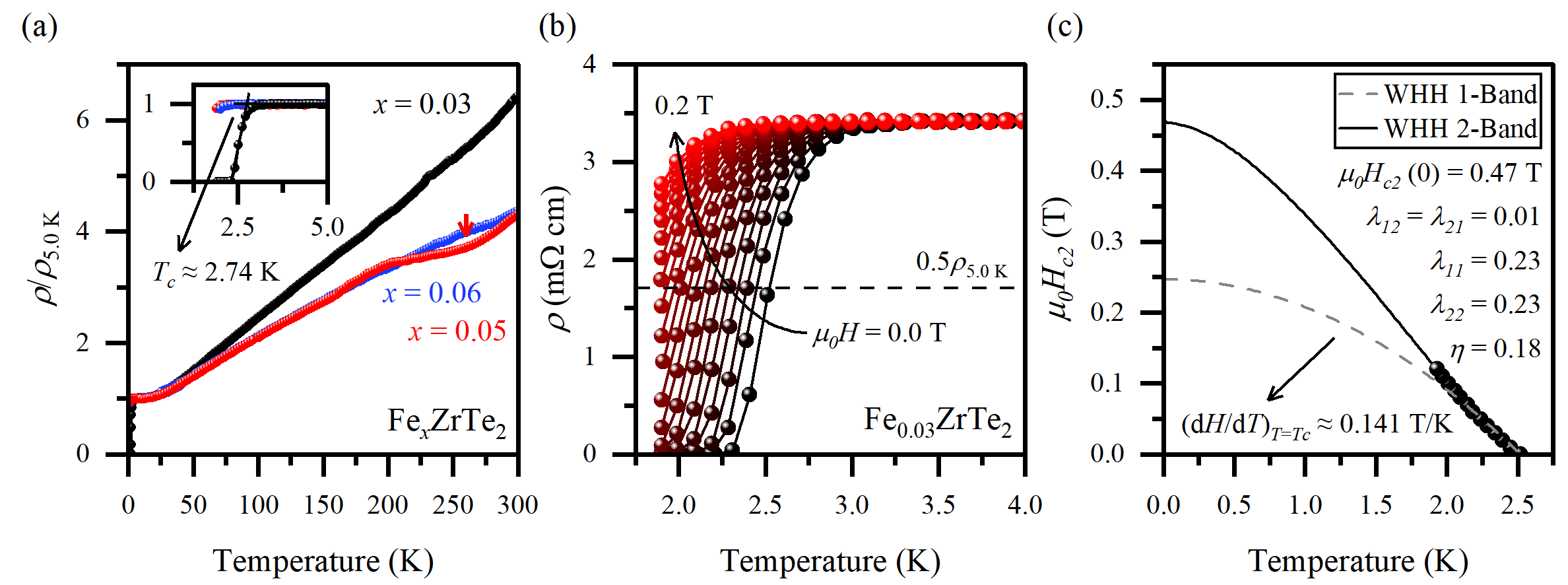}
    \caption{\justifying (a) Electric resistivity $\rho/\rho_{5.0\ \mathrm{K}}$ vs. $T$ for Fe$_x$ZrTe$_2$ single crystals ($x=0.03, 0.05, 0.06$). The inset shows the low temperature $\rho/\rho_{5.0\ \mathrm{K}}$ vs. $T$ with $T_c=2.74$ K for $x=0.03$. Red downward arrow shows the charge density wave transition ($T_{CDW}$). (b) Low temperature magnetorresistivity $\rho$ vs. T for Fe$_{0.03}$ZrTe$_2$ single crystal under magnetic fields up to 0.2 T. (c) Upper critical field $H_{c2}$ vs. T fitted to the WHH single band model (dashed line) and two-band model (solid line) for Fe$_{0.03}$ZrTe$_2$.}
    \label{fig:exp_transport}
\end{figure*}

Fig. \ref{fig:exp_transport}(a) shows the temperature dependence of the electrical resistivity of ZrTe$_2$ intercalated with Fe concentrations ($x$), which for clarity was normalized at 5 K ($\rho / \rho_{5K}$). Throughout the studied temperature range, we observe that $\rho(T)$ decreases with the temperature showing two distinct behaviors: a broad kink-type at high temperatures above 200 K and for concentrations $x \geq$ 0.05, as well as a dropping of $\rho$ towards zero value for all concentrations and at lower temperatures than 4 K. Such behaviors reveal that electron quasiparticles are scattered by two distinct collective quantum orders such as we discuss below.

At low temperatures for $x =  0.03$, the inset of Fig. \ref{fig:exp_transport}(a) shows a drop of $\rho$ below 3.1 K, with the zero resistivity being reached around 2.3 K. This behavior is reminiscent of superconducting (SC) order where the midpoint between these two temperatures determines the critical temperature $T_{c} = 2.74$K (see arrow in the inset of Fig. \ref{fig:exp_transport}(a)). Nearly above $T_{c}$, the quasi-linearity of $\rho(T)$ suggests the presence of metallic behavior. More insights about this superconducting state can be revealed by isofield electrical resistivity curves shown in Fig. \ref{fig:exp_transport}(b), where the superconducting transition is shifted to lower temperatures and broadened with the application of increasing magnetic fields. The upper critical field ($H_{c2}$) at a given temperature was extracted from the 50\% drop of $\rho (T)$ across the transition, which is represented as the dashed horizontal line 0.5 $\rho_{5K}$. Among the available physical scenarios for superconductivity, one can recall the Werthamer-Helfand-Hohenberg (WHH) single-band model which is based on the Ginzburg-Landau (GL) theory of superconductivity. Using this model, The $H_{c2}$ vs. $T$ can be spanned (dashed line in Fig. \ref{fig:exp_transport}(c)) by the following eq.\ref{eq:Hc2WHH} :

\begin{equation}
    ln(t)+U(h)=0
    \label{eq:Hc2WHH}
\end{equation}

where $t=T/T_c$ and $U(h)=\psi(h^*+1/2)-\psi(1/2)$, $\psi(x)$ being the digamma function and $h^*=2H_{c2}/(-\pi^2T dH_{c2}/dT)$. The slope $dH_{c2}/dT \approx 0.141$ T/K was estimated based on the average value in the vicinity of $T_c$ and yielded $\mu_0 H_{c2}\approx0.24$ T.

On the other hand, another scenario for superconductivity that may describe our data considers a two-band model (solid line in Fig. \ref{fig:exp_transport}(c)), proposed by Gurevich, which uses the eq. \ref{eq:Hc2gurevich}:
\begin{equation}
\begin{split}
a_0[\ln(t)+U(h)][\ln(t)+U(\eta h)] + \\
a_1[\ln(t)+U(h)] + a_2[\ln(t)+U(\eta h)] = 0.
\end{split}
\label{eq:Hc2gurevich}
\end{equation}

As we can see in Fig. \ref{fig:exp_transport}(c), the extrapolation to zero temperature yields $H_{c2}(0)=0.47$ T and interband coupling constants $\lambda_{12}=\lambda_{21}=0.01$ significantly lower than intraband coupling constants $
\lambda_{11}=\lambda_{22}=0.23$, suggesting interband Cooper pairing is unlikely. In our present temperature range of measurements, we cannot confirm the sign of the curvature, however, we observe that the temperature dependence of $H_{c2}$ tends towards a positive curvature overall. Therefore, in comparison with a similar multiband scenario for superconducivity associated with ZrTe$_2$ and other chalcogenide systems \cite{correa2022, correa2023, correa2024, Faria2024, ishikura2025superconductivity}, it is appropriate to consider that our $H_{c2} (T)$ data can be better described by eq. \ref{eq:Hc2gurevich}.

Other findings in the electrical resistivity are revealed as the Fe intercalated concentration increases. For Fe$_{0.05}$ZrTe$_{2}$, the onset of superconductivity is shifted to lower temperatures around 2.0 K as seen in Fig. \ref{fig:exp_transport}(a), but no observation of zero resistivity can be distinguished because of the limit in our accessible temperatures ($\sim$ 1.9 K). It is also observed a broad kink centered at $T_{CDW} \approx$ 200 K (see downwards arrow in Fig. \ref{fig:exp_transport}(a)). An accurate determination of $T_{CDW}$ is obtained by the temperature where the first derivative of the electrical resistivity has a minimum (data not shown). In comparison with other TMD systems \cite{machado2017,correa2022,correa2023,correa2024}, the feature observed at $T_{CDW}$ is an indication of the emergence of charge density wave (CDW) order below $T_{CDW}$, which was reported to compete with the superconducting order at low temperatures.

More insights about competition between charge density wave and superconducting orders are inferred from the temperature dependence of electric resistivity of the highest concentration Fe$_{0.06}$ZrTe$_2$ as depicted in Fig. \ref{fig:exp_transport}(a). In comparison with Fe concentration $x =$0.05, the Fe$_{0.06}$ZrTe$_2$ shows similar $T_{c}$ but with markedly increase of $T_{CDW}$ to $\approx$ 265 K. Very interesting, the absence of CDW feature in $x = 0.03$, a composition that shows the highest $T_{c}$, indicates that CDW order becomes weaker as the superconducting order become stronger. Such unusual competition between superconducting and charge density wave electronic states has also been reported in frustrated Kagome lattices \cite{Yu2021}. Likewise frustration, the Fe intercalated substitution plays the role to increase of opening of the gap at the Fermi level ($E_{F}$) responsible for the stabilization of CDW order meanwhile it also reduces the density of states at $E_{F}$, the latter driving to a diminishing of the superconducting critical temperature.

\subsection{Electronic-structure calculations of Fe-doped ZrTe$_2$}

\begin{figure*}[t]
    \centering
    \begin{subfigure}[c]{0.32\linewidth}
        \centering
        \includegraphics[width=\linewidth]{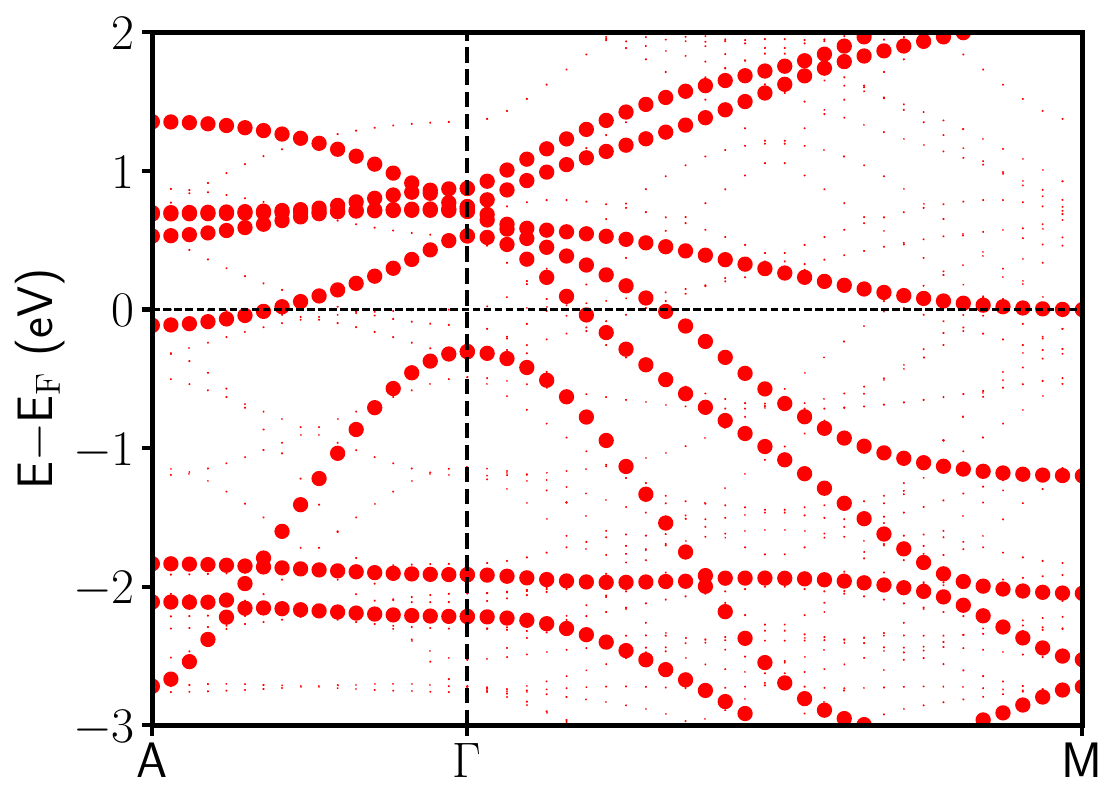}
        \caption{}
        \label{fig:0.00unf}
    \end{subfigure}
    \hfill
    \begin{subfigure}[c]{0.32\linewidth}
        \centering
        \includegraphics[width=\linewidth]{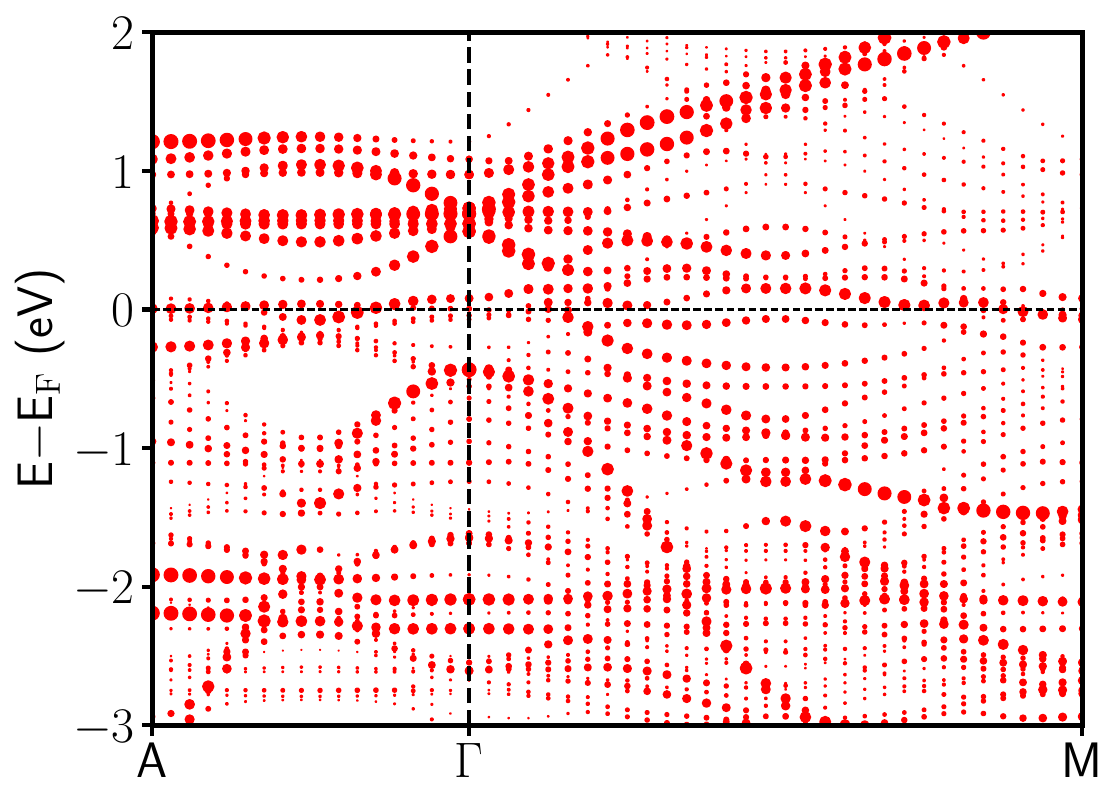}
        \caption{}
        \label{fig:0.125unf}
    \end{subfigure}
    \hfill
    \begin{subfigure}[c]{0.32\linewidth}
        \centering
        \includegraphics[width=\linewidth]{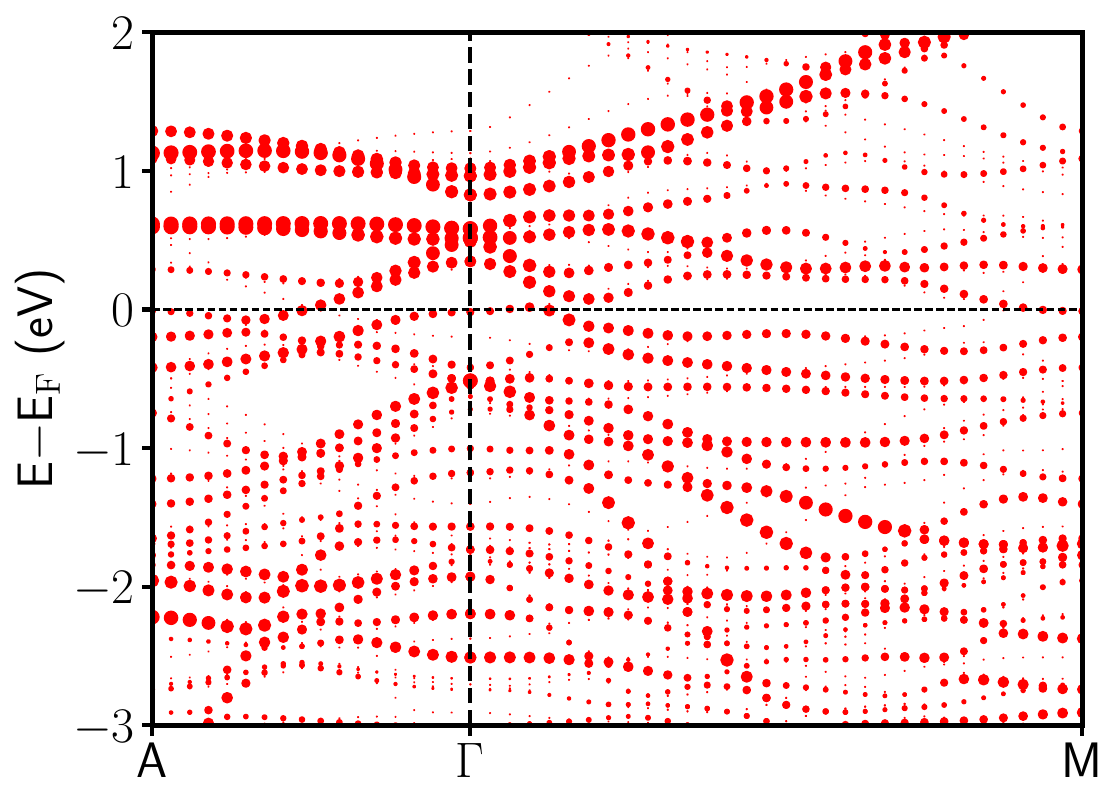}
        \caption{}
        \label{fig:0.25unf}
    \end{subfigure}
    
    \caption{\justifying Unfolded band structures of the $2\times2\times2$ supercells with the following compositions: (a) ZrTe$_2$; (b) Fe$_{0.125}$ZrTe$_2$ (c) Fe$_{0.25}$ZrTe$_2$. The energy range was extended to $-3<E<2$ eV in respect to the Fermi level (E$_f$). $A-\Gamma- M$ path was selected and SOC was taken into account.}
    \label{fig:unfoldings}
\end{figure*}

\begin{figure}
    \includegraphics[width=\linewidth]{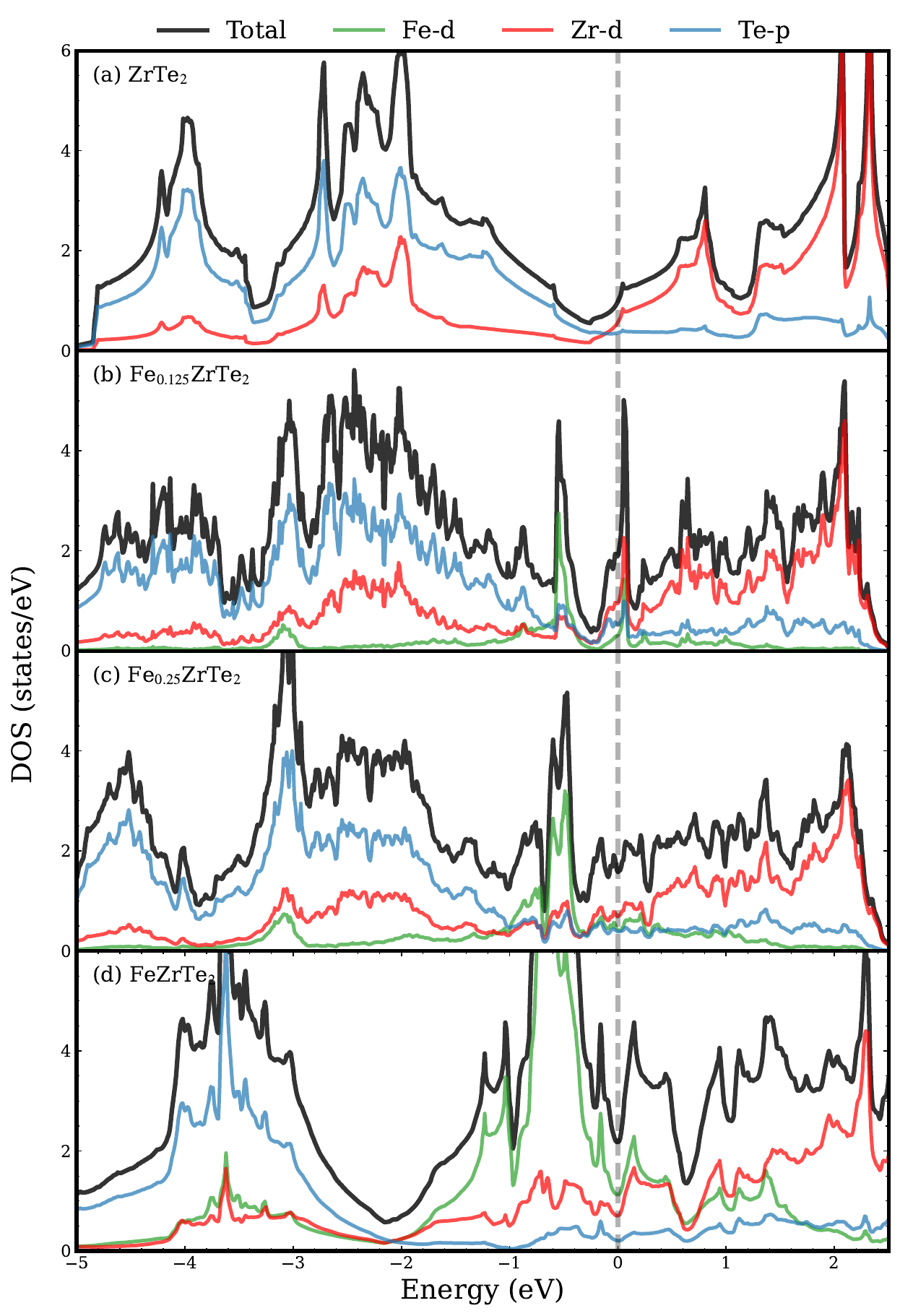}
    \caption{\justifying Projected density of states normalized per unit cell in the range of -5 to 2.5. (a)/(d) are from the stoichiometric compounds with SOC, and (b)/(c) from supercell calculations of $2\times2\times2$ without SOC inclusion.}
    \label{fig:dos4x1}
\end{figure}

For the analysis of the electronic structure of the various possible Fe concentrations, we performed calculations for the primitive cells of ZrTe$_2$ and FeZrTe$_2$, and then, created $2\times2\times2$ supercells of Fe$_x$ZrTe$_2$ ($x=0.0,0.125,0.25$), to achieve more realistic compositions of Fe intercalation in the crystal. In this paragraph, we discuss only the $x=0$ case. The electronic structure of pristine ZrTe$_2$ at the Fermi level (E$_F$) region is constituted by 3 bands (see Figs. \ref{fig:zrte2}(a)-(c) in Appendix \ref{zrte2-app}), with a semimetallic character, where 2 valence bands with a predominant Te-p character have $\Gamma$-centered hole pockets, and a conduction band with Zr-d character with a $L$-centered electron-pocket. Its density of states (DOS), depicted in Fig. \ref{fig:dos4x1}(a), is dominated by Te-p states up to $-0.25\,$eV with respect to $E_F$, where a pseudogap occurs, and Zr-d states start to rise up to a peak at $0.8$ eV. Little hybridization is observed in the region near $E_F$, whereas a considerable hybridization is found for states around $-2.4$ eV.  We further verified its topological character when spin-orbit coupling is taken into account, with a Dirac cone between $\Gamma-A$, visible in Fig. \ref{fig:unfoldings}(a), due to an incompatibility with its irreps, in agreement with previous calculations from \cite{correa2022, https://doi.org/10.1002/advs.202301332}. Other studies reported a nodal line \cite{PhysRevB.102.165149} instead of the fourfold degenerate tilted Dirac cone, and to verify whether this could be a strain or crystal parameters dependence, we additionally calculated 7 more structures: 3 with different fixed lattice parameters from experimental and computational results, but with atomic position relaxation, and other 4 with fixed lattice parameters but changing  $z$ of Te position (see Appendix \ref{zrte2-app} for more details). The result from our PBE-GGA in the 3 first configurations is still a fourfold degenerate point far above E$_f$, but we observed that when the Te position is closer to the Zr planes ($z=0.267\rightarrow z=0.24$), the reported nodal line state occurs, so it would be interesting to perform further studies applying strain on the ZrTe$_2$ cell to change the level of hybridization between Te and Zr states to better characterize this transition. Finally, the three Fermi sheets of pristine ZrTe$_2$ are an enclosed deformed sphere in the $\Gamma-A$ ($c^*$) direction, followed by an unclosed neck-like sheet in the same direction, and those are the 2 valence bands. The last sheet, the conduction band, with low Fermi velocity, is composed by 6 alternating electron pockets alongside $L_{top}-M-L_{bottom}$ direction with connection to  higher order BZs.

We now come to the discussion about the Fe-intercalated Fe$_{x}$ZrTe$_2$ structures with $x=0.125$ and $x=0.25$. Fig. \ref{fig:dos4x1}(b)-(c) presents the DOS, normalized per unit cell. We will describe and analyse the DOS starting from the valence states, far below E$_f$, up to the conduction states. The peak localized at $-3$ eV, is reinforced by low but non-negligible Fe-d orbital contribution. The character of the hybridized descending plateau from around $-2.4$ eV is preserved, with basically the same decreasing trend up to $-1$ eV. Above this energy, a narrow peak appears, formed mostly by Fe-3d localized states at $-0.6$ eV, governed by flatband-like states. The same Te-p/Zr-d pattern happens in both $x=0.125$ and $x=0.25$ cases, but, for $x=0.125$, a new localized state just at the $E_F$ vicinity with Zr-d manifold predominance occurs, increasing the DOS from $1$ (for $x=0$) to a sharp $5$ states/eV. We argue that the stabilized superconducting state that emerges through Fe doping is a consequence of the Fermi surface approaching a van Hove singularity (vHs), as illustrated by the Fe$_{0.125}$ZrTe$_2$ result. The low Fe level causes a perturbation in the the bands slightly above $E_F$ ($\approx54$ meV) in the form of bands with inflection points ($\Gamma-M$ path in Fig. \ref{fig:0.125unf}) \cite{wu2024discovery}. A very similar behavior of this kind of logarithmic instability in the DOS peak giving rise to a superconducting state was already reported for deficient ZrTe$_{1.8}$ \cite{correa2023}, and some other topological/low dimensional materials \cite{wan2023superconducting,luo2023unique,PhysRevB.111.014507,hossain2025superconductivity}. Although this feature is not present in pristine ZrTe$_2$, it is for PdTe$_2$ (vHs above $E_F$ at $M$ point), in which an enhancement of the T$_c$ via Cu doping was attributed to the electron injection pushing the saddle point downwards \cite{PhysRevB.97.165102,cuxpdte2}. Also, recently McFarlane et al identified vHs centered around the same point in the BZ for NiTe$_2$ \cite{hp1t-zd7y}.

We also performed band unfolding of supercell crystals for $x=0,0.125,0.25$, depicted in Fig. \ref{fig:unfoldings}, using the $A-\Gamma-M$ path that was selected for future comparisons with ARPES measurements \cite{Zhang2020Jul}. As a cross-check for the pristine ZrTe$_2$, in Fig. \ref{fig:0.00unf}, we verified the same nature of that reported by Kar et al. \cite{kar2020} for the same path. First, it is possible to verify that, apart from the perturbations introduced in the $x=0.125$ case because of the symmetry reduction, Fe atoms induce an electron doping, with $E_F$ shifted upwards with increasing Fe contents. The Dirac cone located at $0.7$ eV is gradually flattened, and the second conduction and penultimate valence bands, that in the pristine cell were not degenerated, shift towards $0.5$ eV at $A$ point. As mentioned in the above paragraph, The Fe-$3$d states are well localized at $-0.6$ eV and at the near flat-band states at $E_F$ which causes the high peak at the DOS, which are a result of the SOC splitting (see Supplementary Material for the unfolded bands without SOC). In the $x=0.25$ supercell, on the other hand, the bands are less perturbed than in the latter state, as the inversion symmetry is preserved, and an apparent gap opens at $E_F$, and another gap at approximately $0.8$ eV, breaking the touching of degenerate bands at $\Gamma$. We should also bear in mind that our supercell calculations do not took into account all possible symmetry nonequivalent sites at $x=0.25$. Instead, the Fe atoms were randomly placed at the $1b$ positions.

\begin{figure}
    \centering
    \includegraphics[width=\linewidth]{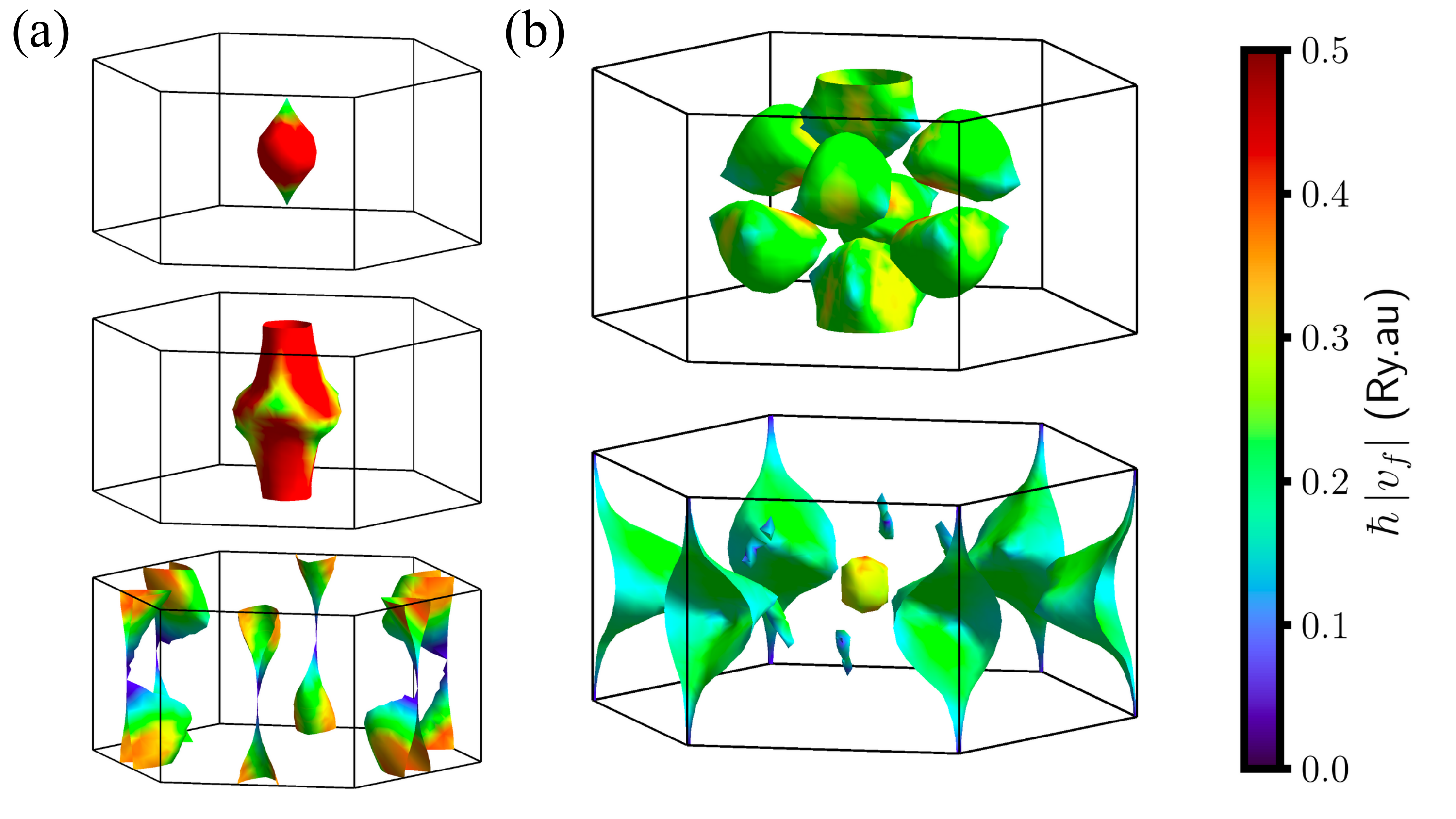}
    \caption{\justifying The independent sheets of the Fermi surface of (a) pristine ZrTe$_2$ and (b) intercalated FeZrTe$_2$. The color map indicates the Fermi velocity magnitude in natural units ($1$ Ry.au $\approx2.187\times10^6$ m/s).}
    \label{fig:fermi}
\end{figure}

\begin{figure*}[t]
    \includegraphics[width=\linewidth]{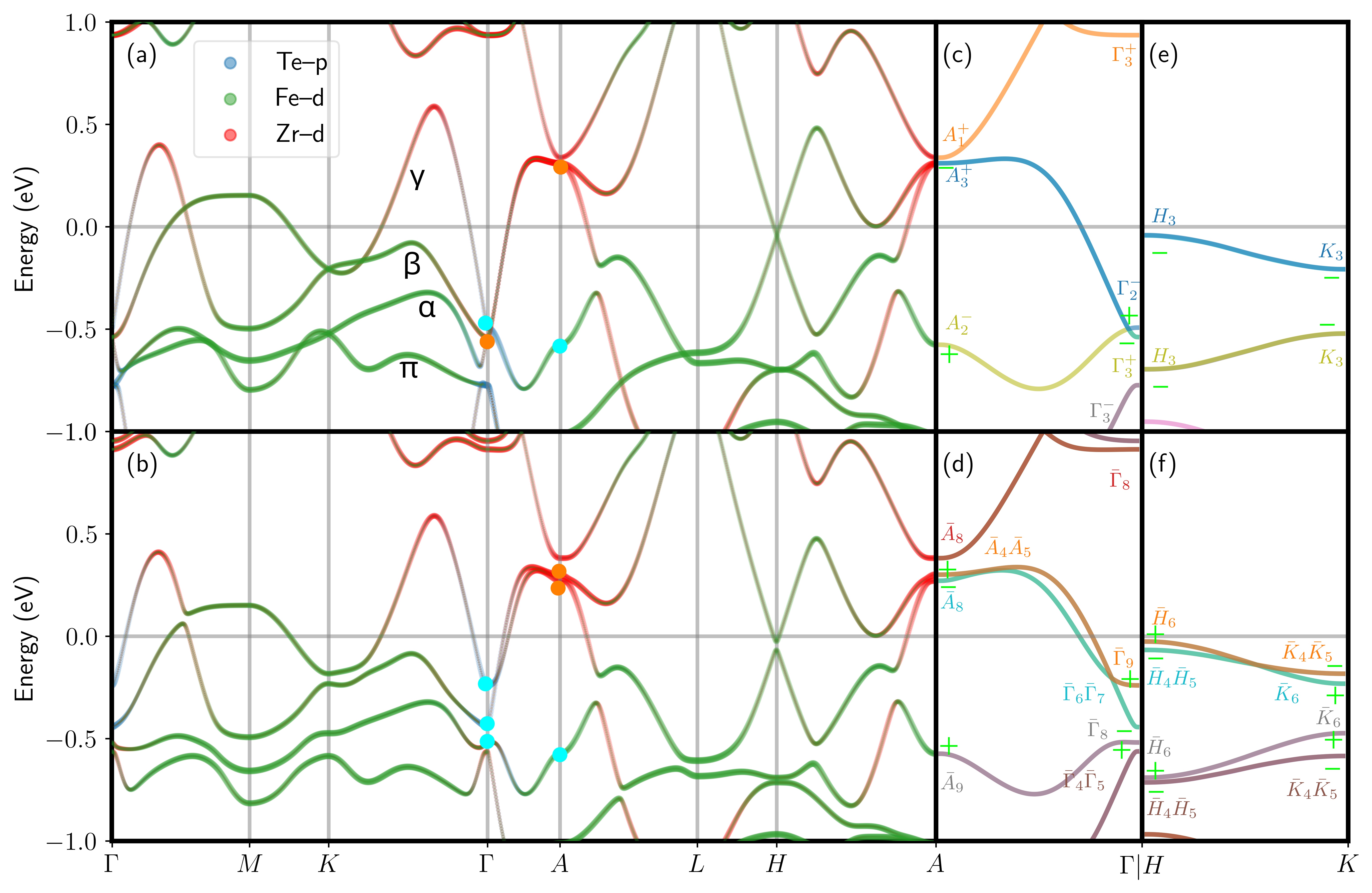}
    \caption{\justifying Band structure of stoichiometric FeZrTe$_2$. In the first (second) line SOC was not considered (included). (a)-(b) A complete plot of the eletronic dispersion along high-symmetry points of the BZ. (c)-(f) Plots of selected paths where the band crossing occurs, with different colors assigned to each band and their respective irreducible representations. Orange and blue points represent $+/-$ inversion eigenvalues, and green $+/-$ symbols represent the $\{3^{\pm}_{001}|0\}$ eigenvalues.}
    \label{fig:bands-irreps}
\end{figure*}

In the stoichiometric ideal compound, FeZrTe$_2$ ($x=1$), Fig. \ref{fig:dos4x1}(d), the electronic structure is completely changed. There is a deep valley in the DOS around $-2$ eV, instead of the several peaks found for the other compositions, including pristine ZrTe$_2$. A change in the major contributor from Te to Zr happens at $-2$ eV, and at $-3.6$ eV a peak with strong hybridization between Fe/Zr/Te occurs, where Fe-d and Zr-d contributes equally to the DOS, but after the valley, Fe-d states rise and dominates the DOS up to and above $E_F$, starting from the large localized peak at $-0.6$ eV, the same energy level as in the previous supercells, but with a much larger DOS ($6$ states/eV). 
The electronic structure at the region near $E_F$, evidenced in Fig. \ref{fig:bands-irreps} is constituted by two bands crossing the Fermi level: a valence band with Fe-3d character predominance, that has a tiny hole pocket along $\Gamma-M$ and another centered in $A$, which gives rise to three dimensional 6 enclosed low-velocity quasi-spherical sheets with Fe-3d predominance and 2 pockets in $A$ with Zr-4d predominant Fermi sheets, respectively, in Fig. \ref{fig:fermi}(b); and a conduction band with electron-like character that has a $\Gamma$-centered steep parabolic-like electron pocket, with Te-p character, a less dispersive $K$ and a tiny $H$ pocket with Fe-3d character, that generates a Fermi surface with six hat-like pointing from $K$ to $\Gamma$, a spherical sheet with high Fermi velocity in $\Gamma$ surrounded by six droplets in $\Gamma-L$ direction (cf. Fig. \ref{fig:fermi}). 

The spin-orbit coupling (SOC) effects break degeneracies, in particular, bands $\gamma$ and $\beta$, in the case without SOC, that have degeneracies between $\Gamma-M$ (fourfold point) and $K-\Gamma$; a fourfold degenerate line between $H-K$ where the electronic eigenfunctions transform as two dimensional irreps $H_3$ and $K_3$; in the $\Gamma-A$ path, where bands $\alpha$ and $\beta$ are degenerated in $\Gamma$ with irrep $\Gamma^+_3$ and band $\gamma$ has irrep $\Gamma_2^-$. In $A$ point, the irreps of bands $\beta$ and $\gamma$ transform into $A_3^+$, and band $\alpha$ irrep into $A^-_2$. Thus, band $\beta$ is degenerate along the entire path, but with band $\alpha$ up to the sixfold crossing point, and after it with $\gamma$. By extracting the inversion eigenvalues ($+$ as orange and $-$ as blue dots in Figs. \ref{fig:bands-irreps}(a)-(b)), a symmetry that is present along $\Gamma-A$, but not along $H-K$, we also verified that bands $\alpha$ and $\gamma$ are inverted through the path $\Gamma-A$, while band $\beta$ remained with its positive eigenvalue along that line. 

With all bands becoming twofold degenerate with SOC in every $k$-point in Kramers pairs forced by the preservation of time-reversal ($\mathcal{T}$) and inversion symmetries ($\mathcal{P}$), most of the fourfold degeneracies are broken. In the $\Gamma-A$ path, the degeneracy between $\alpha$ and $\beta$ is broken at $\Gamma$, with $\alpha$ having its double-valued irrep $\bar \Gamma_8$ subduced to $\bar\Lambda_9$. Therefore, the dispersion along $A-\Gamma$, besides the broken degeneracy, exhibits also more subtle changes with respect to the non-SOC case.
The other two bands, $\beta$ (last filled band) and $\gamma$, with irreps $\bar\Gamma_6\oplus\bar\Gamma_7,\bar\Gamma_9$ at $\Gamma$ subduced to $\bar\Lambda_8,\bar\Lambda_4 \oplus\bar\Lambda_5$, are splitted. Particularly, $\gamma$ is raised in energy at $\Gamma$. Furthermore, by separating the symmetry operation $\{3^{\pm}_{001}|0\}$ eigenvalues, depicted in green $+/-$ symbols in the selected paths, we verify that in the two paths, band crossings between $\beta$ and $\gamma$ are protected by this rotation symmetry, leading to two type $II$-like Dirac cones, required by the different representations as band hybridization is not allowed. Specially along $H-K$, when SOC is incorporated, the fourfold degenerate bands are broken: $\pi$ and $\alpha$ bands are separated with $-$ and $+$ eigenvalues, respectively, and their irreps are compatible ($\bar\Gamma_4\oplus\bar\Gamma_5$ to $\bar\Lambda_4\oplus\bar\Lambda_5$ and $\bar\Gamma_6$ to $\bar\Lambda_6$). The other bands, $\beta$ and $\gamma$, have the same irreps but they are inverted along the path. Also, separating the Kramers pairs with respect to their inversion eigenvalues, the band inversion that occurred when SOC was not considered is broken as bands $\alpha$ and $\beta$ are splitted and a transition from $+$ to $-$ eigenvalue is observed. This transition causes band $\alpha$ to be in a normal state, with main contribution from Zr-d, but now band $\beta$ is inverted along the path, and its Zr-d character is apparently exchanged to $\gamma$ Te-p character at $\Gamma$ (see Fig. S1 of Supp. Material), supporting the fact that this semimetal has a nontrivial topology \cite{yang2014classification}.

Using \textsc{IrRep}, the traces of every symmetry operators of the double space groups in the maximal $\mathbf k$-vectors of ZrTe$_2$ and FeZrTe$_2$ were computed up to a filling number (in this case, the number of electrons in the valence of the pseudopotential). Also, the Check Topological Mat. tool was used to compare these traces computed within DFT with the BCS character tables. The calculated results indicate that both of them are classified as enforced semimetals (ES), in which the irreps at different high-symmetry points that will necessarily be connected with irreps of bands after the filling. Thus, an isolated filling is impossible \cite{TQC}. So, through intercalating Fe in the pristine compound, which already is an ES with a single connection between valence and conduction bands far above $E_F$, becames another ES, at $x=1$, with two connections below and much nearer to $E_F$, facilitating further confirmation with ARPES experiments, for example.

\section{Conclusion}
Results obtained by means of electric transport measurements in Fe$_x$ZrTe$_2$ single crystals support the existence of a new superconducting state with $T_c \approx 2.74\,$K in chemically doped Fe$_{0.03}$ZrTe$_2$, with the temperature dependence of $H_{c2}$ being fit with a two-band model. Superconductivity in Fe$_x$ZrTe$_2$ seems to become weaker as the the charge density wave electronic state becomes stronger with varying $x$ concentration. A tentative scenario for this unusual competition recalls a mechanism likewise it is found in other systems where frustration is relevant. The intercalated Fe in the crystal may occupy distinct positions that resembles frustration in such a way that the increase of gap opening in the density of state strengthens the CDW order (here, $T_{CDW}=265\,$K in Fe$_{0.05}$ZrTe$_2$) while it diminishes the SC state.
Our experimental findings in the presence of Fe intercalation is supported by our ab initio calculations, in which disorder effects are considered in the electronic band structure. 
Through calculations, we explored Fe$_{0.125}$ZrTe$_2$ and Fe$_{0.25}$ZrTe$_2$, where the density of states and unfolded band structure for the lower Fe percentage in the structure exhibited a van Hove singularity in the vicinity of the Fermi level, which vanishes completely when raising $x$. As stated before, increasing the density of states, due to the flat/localized states, will favor the superconducting order. Furthermore, by comparing the differences in electron localization function, we verified that even with few Fe atoms the change in the local bonding environment causes a disruption of the periodic charge distribution, which ultimately leads to the destabilization of the CDW order. The electronic structure evolution from ZrTe$_2$ to the fully intercalated compound FeZrTe$_2$ was also discussed by calculatig their irreps and traces, in which the enforced semimetallic character is mantained, but with crossings lowered in energies due to the electron injection. Now, these topological features are just under the Fermi level, instead of the studied $\Gamma-A$ Dirac crossing/nodal line far above it. This observation establishes Fe$_x$ZrTe$_2$ as an interesting candidate for studying the coexistence of nontrivial band structure and superconductivity in future works.

\section*{Data availability}
All the relevant computational data of this research is provided in this \href{https://github.com/cauaschuch/Fe_xZrTe_2-DSM}{GitHub repository} https://github.com/cauaschuch/Fe\_xZrTe\_2-DSM and after the paper acceptance will be open source available at Zenodo, also its Supplementary Material.
Experimental data will be made available on request.

\begin{acknowledgments}

We gratefully acknowledge the financial support of the São Paulo Research Foundation (FAPESP) under Grants 2024/21634-2, 2024/23535-1, 2024/01514-2. J. L. J. and A.F.R acknowledge Grants 2018/08845-3, 2022/14202-3, and 2020/01377-4 FAPESP. J. L. J acknowledges Centro Nacional de Desenvolvimento Científico e Tecnológico (CNPq) Grant 308825/2025-0. L.T.F.E acknowledges CNPq Grant 311756/2022-0. The research was carried out using high-performance computing resources made available by the Superintendência de Tecnologia da Informação (STI), Universidade de São Paulo.

\end{acknowledgments}

\appendix

\section{ZrTe$_2$ transition from nodal line semimetal to Dirac semimetal}\label{zrte2-app}
\begin{figure*}[t]
    \centering
    \includegraphics[width=\linewidth]{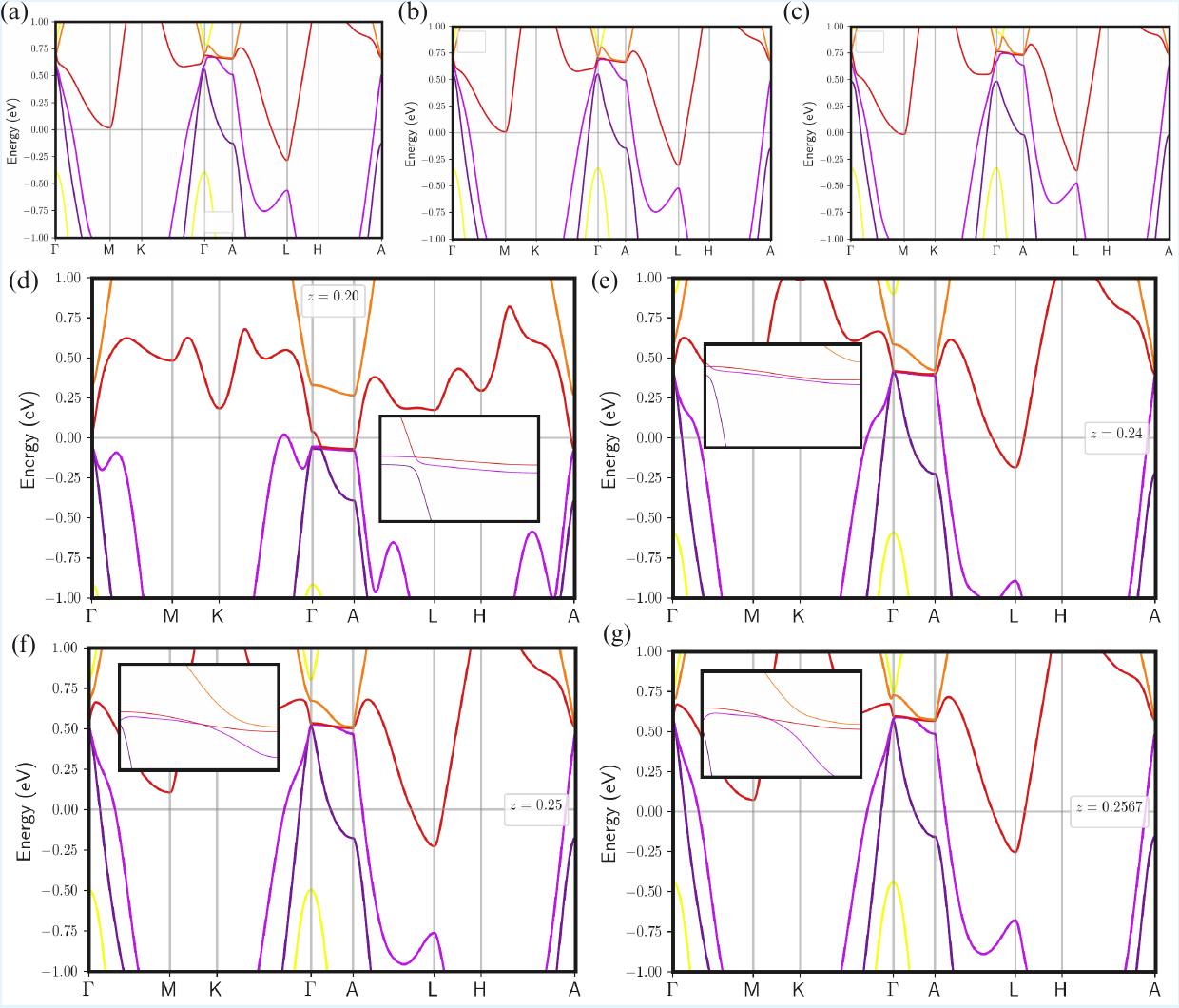}
    \caption{\justifying Band structures of pristine ZrTe$_2$ in other unit cell configurations for comparison: (a) synchrotron
X-ray powder diffraction experiment \cite{shkvarina2018}, $a=3.965,c=6.745$ \AA; (b) Entry $653213$ from the Inorganic Crystalline Structure Database (ICSD) \cite{mctaggart1958sulphides}, which was used in Topological Materials Database search \cite{catalogue-topo,TQC,doi:10.1126/science.abg9094}, $a=3.952,c=6.660$ \AA; (c) DFT from Kar et al.  \cite{kar2020}, $a=3.909,c=6.749$ \AA. The other four plots are using ZrTe$_2$ fixed lattice of (b) but changing vertical Te-$z$ equivalent position of $2d$ Wyckoff site $(\frac{1}{3},\frac{2}{3},z)$: (d) $z=0.2$; (e) $z=0.24 $; (f) $z=0.25$ and (g) $z=0.2567$. Inset zoom was applied into the crossing region between $\Gamma-A$ for (d)-(g) plots. }
    \label{fig:zrte2}
\end{figure*}

To better understand the true fundamental state of ZrTe$_2$, we performed extra DFT calculations of the pristine cell, that are evidenced in Fig. \ref{fig:zrte2}, all with ab initio SOC inclusion and then projected into a pseudo atomic orbital base (PAO) to construct tight-binding hamiltonians to get a better flexibility calculating the electronic eigenvalues along the path, using PAOFLOW code \cite{BUONGIORNONARDELLI2018462PAOFLOW,cerasoli2021advancedmodelingmaterialspaoflow}. From the first three configurations, (a)-(c), in which we relaxed atomic positions (using \textsc{vdW--DF3}), Te-$z$ coordinate was in the range of $0.266-0.269$, and both electronic structures showed the type-II Dirac crossing between the soft purple and red bands, along $\Gamma-A$ line. The degeneracy present in DFT calculations between the soft and dark purple bands present in DFT calculations from Tian et al. \cite{PhysRevB.102.165149} at $\Gamma$ was not achieved, but the orange and red bands touch at $A$ was. So, we selected lattice parameters from (b), that has $z=0.2697$, and manually reduced this value. Reducing up to $z=0.25$, in Fig. \ref{fig:zrte2}(f)-(g), the sixfold degeneracy at $\Gamma$ appeared, with orange band lifted, but the nodal line just emerged with a fourfold degeneracy when $z$ was decreased up to $0.24$. Reducing below $z=0.24$ brings the already splitted nodal line to below E$_f$, in which at an extreme coordinate $z=0.20$ the orange band has no degeneracy in the path, and the tilted Dirac cone returns, approaching the $\Gamma$ point. So, by this analysis, it is clear that the bands stick together below a certain $z$, and that above this value, the line degeneracy will be lifted and the touching point will prevail.

\section{Electron localization function analysis}
\begin{figure*}[t]
    \centering
    \includegraphics[width=\linewidth]{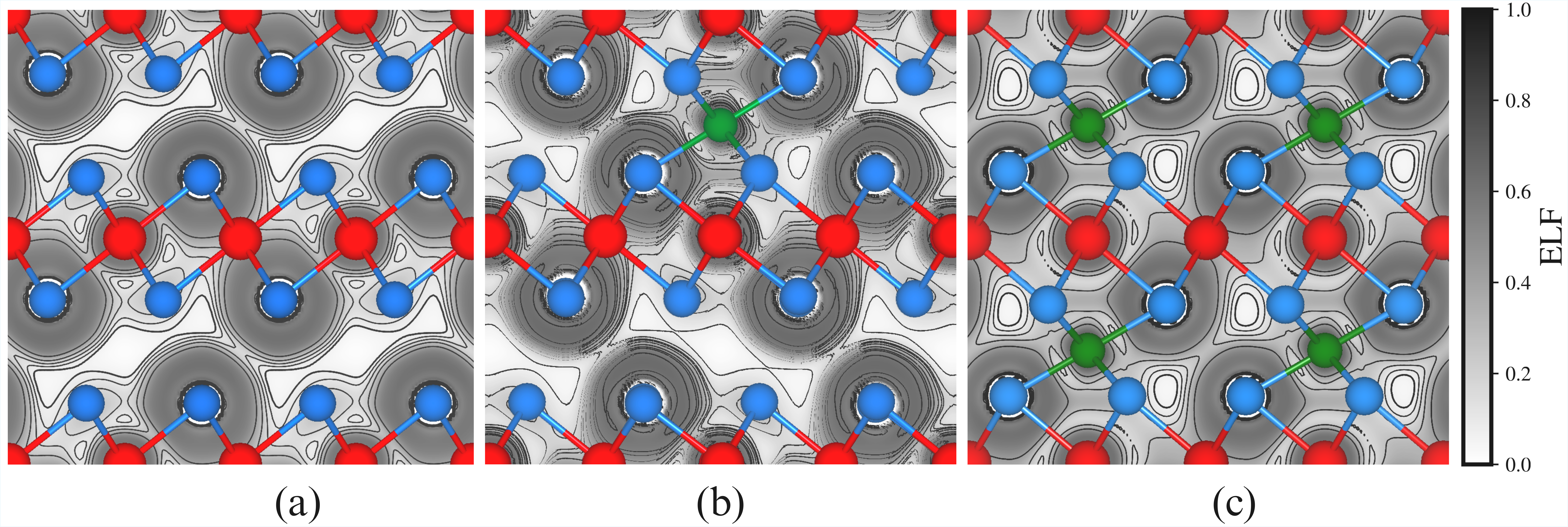}
    \caption{\justifying ELF plotted in the plane $(hkl)=(110)$ for: (a) cell of pristine ZrTe$_2$ , (b) cell of Fe$_{0.125}$ZrTe$_2$ and (c) cell of FeZrTe$_2$.}
    \label{fig:elf}
\end{figure*}

\begin{figure}
    \centering
    \includegraphics[width=\linewidth]{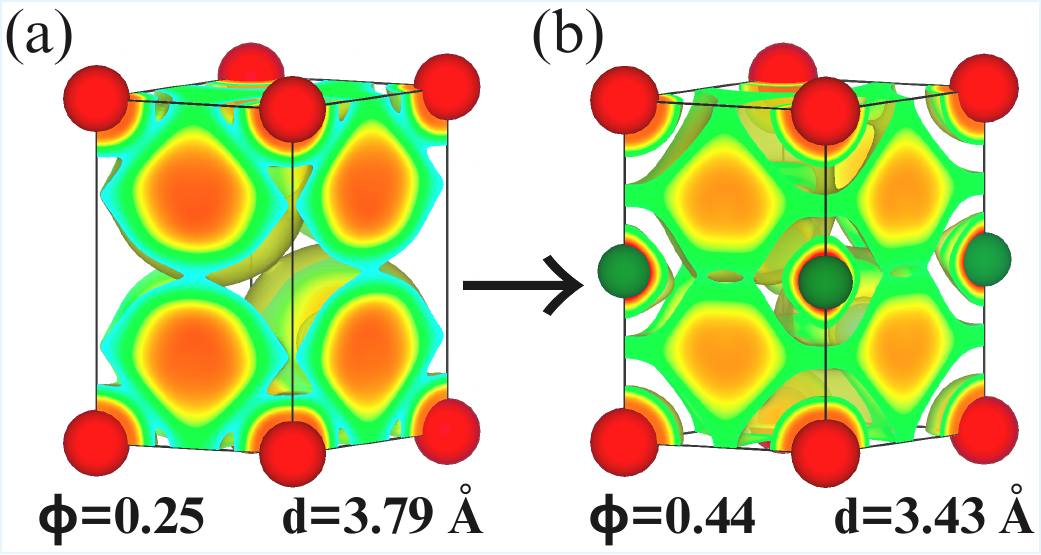}
    \caption{\justifying A 3D visualization of the ELF isosurface: (a) cell of ZrTe$_2$ and (b) cell of FeZrTe$_2$. The $\phi$ and $d$ values are indicatives of the networking value for Te atoms span and Te-Te bonding distance (\AA), respectively.}
    \label{fig:connect}
\end{figure}

The electron localization function (ELF) was employed to gain insights into the different bonding environments of Fe-doped ZrTe$_2$. It can simply be enunciated as ELF $= (1+\chi_\sigma^ 2)^{-1}$, with $\chi_\sigma=D_\sigma/D_\sigma^0$, where $D_\sigma^0$ is a measure of the uniform electron gas electron localization \cite{elf-paper}. Thus, by definition, its range is $0\leq$ ELF$ \leq1$, where a value from $0.8$ to $1$ indicates a paired covalent bond as the system is more localized, and a value less or equal to $0.5$ is the indicative of a delocalized, free electron gas bond \cite{elf1}. In Fig. \ref{fig:elf}, that shows the evolution of ELF in ZrTe$_2$ crystal, we verify that the disorder produced by a single Fe atom (in $x=0.125$) is that of inducing delocalization between layers gap, connecting Te-Te zig-zag interactions that have ELF with value near $0.4$, as the $a-b$ planes interlayer spacing is reduced. Furthermore, in the stoichiometric compound ($x=1)$, the isovalues near $0$ are basically vanished, as Fe closes completely the van der Waals gap, and the metallic character has more dominance. The connectivity analysis was based in the theory developed by Belli et al. \cite{belli2021strong}, that is centered in finding the ELF value for which its isosurface spans the entire unit cell of the compound, surpassing the atomic region and creating a bridge between atoms. The connectivity value $\phi$ was determined and depicted in \ref{fig:connect}. With this, we verified that through Fe doping $\phi$ is increased from $\phi=0.25$, in the pristine compound, to $\phi=0.28$ in $x=0.125$ supercell up to the limit of $\phi=0.44$, as the distance $d$ of Te atoms is reduced. Following the model of real space representation of the Cooper pair wave function of \cite{belli2021strong}, increasing the $\phi$ value will increase the SC state delocalization, leading to an $e$--ph coupling enhancement. Although developed in the context of hydrogen-based superconductors, the approach produces results for our compound that follow the same trends reported in the original work.
%\nocite{*}

\bibliography{apssamp}% Produces the bibliography via BibTeX.

\end{document}